\documentclass[pre,twocolumn,floatfix]{revtex4-2} 

\usepackage{amsmath} 
\usepackage{amsfonts} 
\usepackage{graphicx} 

\begin{document}


\title{Intermittency as a consequence of a stationarity constraint on the energy flux}
\author{S\'ebastien Auma\^ \i tre,}
\email[Corresponding author. Email address: ]{sebastien.Aumaitre@cea.fr}
\affiliation{SPEC, CEA, CNRS, Université Paris-Saclay, CEA Saclay, 91191, Gif sur Yvette, Cedex, France}
\author{St\'ephan Fauve}
\affiliation{Laboratoire de Physique de l'\'{E}cole Normale Sup\'{e}rieure, CNRS, PSL Research University, Sorbonne Universit\'{e}, Universit\'{e} de Paris, F-75005 Paris, France }

\begin{abstract}
{ In his seminal work on turbulence, Kolmogorov made use of the stationary hypothesis to determine the Power Density Spectrum of  the velocity field in turbulent flows. However to our knowledge, the constraints that stationary processes impose on the fluctuations of the energy flux have never been used in the context of turbulence. Here we recall that the Power Density Spectra of the fluctuations of the injected power, the dissipated power and the energy flux have to converge to a common value at vanishing frequency. Hence, we show that the intermittent GOY--shell model fulfills these constraints. We argue that they can be related to intermittency. Indeed, we find that the constraint on the fluctuations of the energy flux implies a relation between the scaling exponents that characterize intermittency, which is verified by the GOY--shell model and in agreement with the She-Leveque formula. It also fixes the intermittency parameter of the log-normal model at a realistic value. The relevance of these results for real turbulence is drawn in the concluding remarks. }\\
\end{abstract}
\maketitle


The complex structure and the statistics of turbulent flows still resist to a full understanding. The large number of spatial scales between the injection and the dissipation (efficient only at very small scale) also implies many time scales. This makes full numerical studies of temporal fluctuations in the limit of very turbulent flow (at high Reynolds number) very costly. The intermittency of the velocity increments with their highly non-Gaussian fluctuations has been widely discussed, although the mechanisms implying such complexity remain unclear in turbulent flows \cite{Frish}.
In 1941, Kolmogorov (K41 theory) used the ingredients of  stationarity,  locality of the nonlinear interactions and  existence of a large inertial range of scales where the energy injected at large scale is transferred lossless down to the small dissipative scales. Stationarity imposes that the rate of energy transfer, called the {\it energy flux}, is equal to the injected power on average. Moreover, for homogeneous isotropic turbulence, the velocity increments on size $l$, $\delta_lu(r)= \left( u(r+l)-u(r)\right)$, is assumed to depend only on the mean dissipated power per unit mass, $\langle D \rangle$, and on  $l$ in  the inertial range. All together this implies that the third moment of the velocity increment, $\langle \delta_lu(r)^3\rangle$, is proportional to $\langle D \rangle l$. Because it does not take into account the fluctuations in the process of energy transfer, K41 theory fails to correctly predict the scaling exponents $\zeta(p)$ defined by $\langle \delta_lu(r)^p\rangle\propto l^{\zeta(p)}$ for $p\neq 3$. Based on the same stationarity hypothesis, many refinements of the theory tried to obtain these exponents \cite{Frish,Yaglom,K62,Benzi84,Meneveau87,Kida90,SLmodel}. They take into account the fluctuations of the dissipation rate that depend on the complex structure of the flow in order to get the exponents $\zeta(p)$ in agreement with experiments \cite{AnselmetGagneHoppfinger,ESS}. However, the underlying requirement that imposes such complexity remains hidden.   

She and Leveque proposed one of these models. They predict \cite{SLmodel}:
\begin{equation}
\zeta(p)=\gamma p+(1-3\gamma)\frac{\left(1-\beta^{p/3}\right)}{1-\beta}
\label{SheLeveque}
\end{equation}
This is in very good agreement with the measurements of turbulent flows for $\gamma=1/8$ and $\beta=0.58$ \cite{SLmodel}. We are going to use it as a reference. 
We will also consider the log-normal model that gives another function for the scaling exponents\cite{Yaglom,K62}
\begin{equation}
\zeta(p)=p/3+(\mu/18)(3p-p^2),
\label{LogNormalGOY}
\end{equation}
which involves a single free {\it intermittency parameter} $\mu$. 

It is remarkable that intermittency is observed in models of turbulent flows known as shell models in which geometrical structures of the flow are discarded and the Kolmogorov cascade is described by a set of ordinary differential equations for velocities $u_n(t)$ related to the energy content of scales $k_n \propto 2^n$\cite{OY,Kadanoff}. The scaling exponents $\zeta(p)$ defined above in real space and defined here by
\begin{equation} 
\langle |u_n|^p\rangle\sim k_n^{-\zeta(p)}.
\label{scaling}
\end{equation}
are well described by (\ref{SheLeveque}) \cite{SLPRE}.

This quantitative agreement for scaling exponents measured in turbulent flows and obtained in shell models motivated us to find a general constraint on turbulent cascades of energy, independent of the flow geometry, that could explain intermittency corrections to K41 theory. To wit, we use an additional constraint imposed by stationarity beyond the equality of the averages of injected power, energy flux and dissipation. This additional constraint is related to the Power Density Spectra (PDS) of injected and dissipated power, that should be equal in the limit of vanishing frequency.
We show that the intermittent GOY--shell model indeed satisfies this constraint on power. Furthermore, the same constraint is also fulfilled by the energy flux in this case. This compels the product of the variance of the energy flux with its correlation time to be constant within the inertial range. We then discuss how this constraint imposes a relation between two scaling exponents $\zeta(p)$ for different $p$'s. We check that our simulation of the GOY-shell model as well as  the She and Leveque (SL) prediction \cite{SLmodel,SLPRE} satisfies this relation. We also demonstrate that this relation fixes the intermittent parameter of the log-normal model to its expected value \cite{Yaglom,K62}. We finally discuss the relevance of these results for real 3D turbulence.\\


Turbulent flows like all dissipative systems in a stationary state obeys the energy balance 
\begin{equation}
\frac{dE}{dt}=I-D,
\label{EBalance}
\end{equation} 
where $E$ is the total internal energy, $I$ is the injected power and $D$ the dissipated power \cite{Aumaitre2001}. Obviously properties of stationary processes impose $\langle I \rangle=\langle D \rangle$ where $\langle~\cdot~ \rangle$ stands for the ensemble average. One can go further by taking the Fourier transform of equation (\ref{EBalance}) and multiplying it by $\left(\widehat{I}(\omega)+\widehat{D}(\omega)\right)^*$ where $\widehat{X}$ is the Fourier transform of the centered variable $\Delta X=X-\langle X \rangle$ and $^*$ stands for complex conjugate. Then one takes the limit $\omega\rightarrow 0$ on the real part and obtains a second order stationary relation
\begin{equation}
\lim_{\omega\rightarrow 0} \left[|\widehat{I}(\omega)|^2\right]= \lim_{\omega\rightarrow 0} \left [|\widehat{D}(\omega)|^2\right].
\label{PDSstat}
\end{equation}
Using the Wiener-Kinchine theorem, one can write (\ref{PDSstat}) as
\begin{equation}
\int_0^\infty\langle\Delta I(t)\Delta I(t+\tau)\rangle d\tau=\int_0^\infty\langle\Delta D(t)\Delta D(t+\tau)\rangle d\tau.
\label{WKint}
\end{equation}
An alternative demonstration of this equality was first proposed by J. Farago in \cite{Farago}.
In order to perform scaling arguments, it is convenient to introduce the correlation time of the variable $X$ defined as $\tau_X=\frac{1} {\sigma(X)^2}\; \int_0^{+\infty} \langle \Delta X(t)\; \Delta X(t+\tau) \rangle d\tau $. With this definition, one can rewrite equation (\ref{WKint}) as:
\begin{equation}
\sigma(I)^2\tau_I=\sigma(D)^2\tau_D
\label{BalanceII}
\end{equation}
The pertinence of this equation have been tested on various dissipative systems in \cite{Aumaitre2004,AumaitreNaertAppfel}. Equation (\ref{BalanceII}) relates the low frequency fluctuations of injected and dissipated power that are usually not considered in turbulent flows. Moreover, in the case of turbulent shell models, it is possible to extend the balance (\ref{EBalance}) to the energy flux, as we will show hereafter. In addition, such simplified models are convenient to study the asymptotic behaviors at vanishing frequency because they are easy to integrate numerically over very long time. We will focus on the GOY--shell model in the following \cite{OY}.\\


The GOY--shell model is built to exhibit a highly intermittent behavior. In our computations, we discretize the wavenumber space in $N=20$ shells of size increasing exponentially. The complex velocity $u_n=u_{n,r}+j\cdot u_{n,i}$ in the $n^{th}$ shell is described by the set of equations \cite{Frish,SLPRE,OY,Kadanoff}:

\begin{equation}
\frac{d u_n}{dt}={\mathcal F}(u_{n-2},u_{n-1},u_{n+1},u_{n+2})+f_4\cdot \delta_{n,4}-\nu k_n^2 u_n
\label{GOYdyn1}
\end{equation}

\noindent
where $k_n=2^n/16$ is the wave number corresponding to the shell $n$ ($1 \leq n \leq N$), $\nu$ is the kinematic viscosity ($\nu= 7.62\times 10^{-7}$ in order to keep few shells in the dissipative range) and $f_0$ represents a constant forcing acting on the 4th shell only. The large forcing scale is thus $L=k_4^{-1}$ and has a characteristc velocity $U=\sqrt{f_o/k_4}$. One chooses the nonlinear terms in order to conserve the volume in phase space, energy and helicity in the inviscid limit i.e. ${\mathcal F}(u_{n-2},u_{n-1},u_{n+1},u_{n+2})=jk _n(u_{n+1} u_{n+2}-u_{n-1} u_{n+1}/4-u_{n-1} u_{n-2}/8)^*$ \cite{OY}. 

\begin{figure}
\vspace{-3cm}
\resizebox{.5\textwidth}{!}{%
\includegraphics{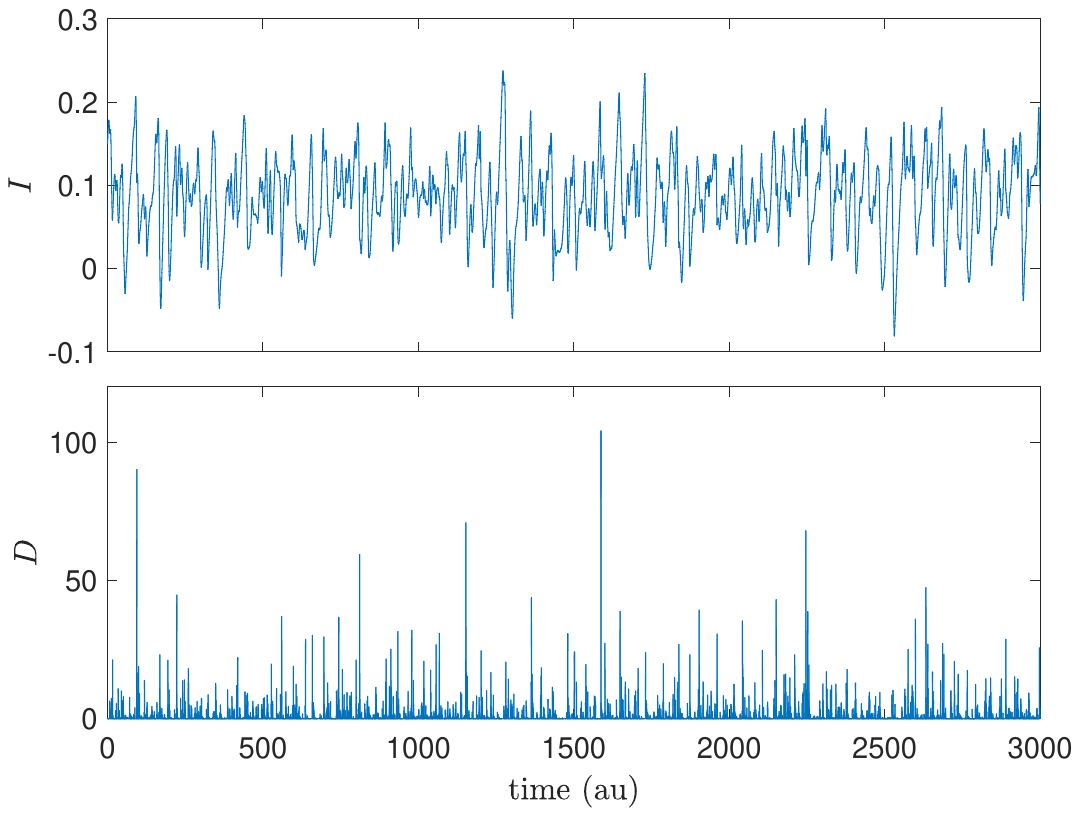}
}
\vspace{-4cm}
\caption{Temporal trace of the injected (top) and dissipated (bottom) power in the GOY--shell model with N=20, $\nu= 7.62 \, 10^{-7}$ and $f_0=0.071$}
\label{TrcTpesIDGOY}
\end{figure}

\begin{figure}
\vspace{-3cm}
\resizebox{.5\textwidth}{!}{%
\hspace{-0.5cm}
\includegraphics{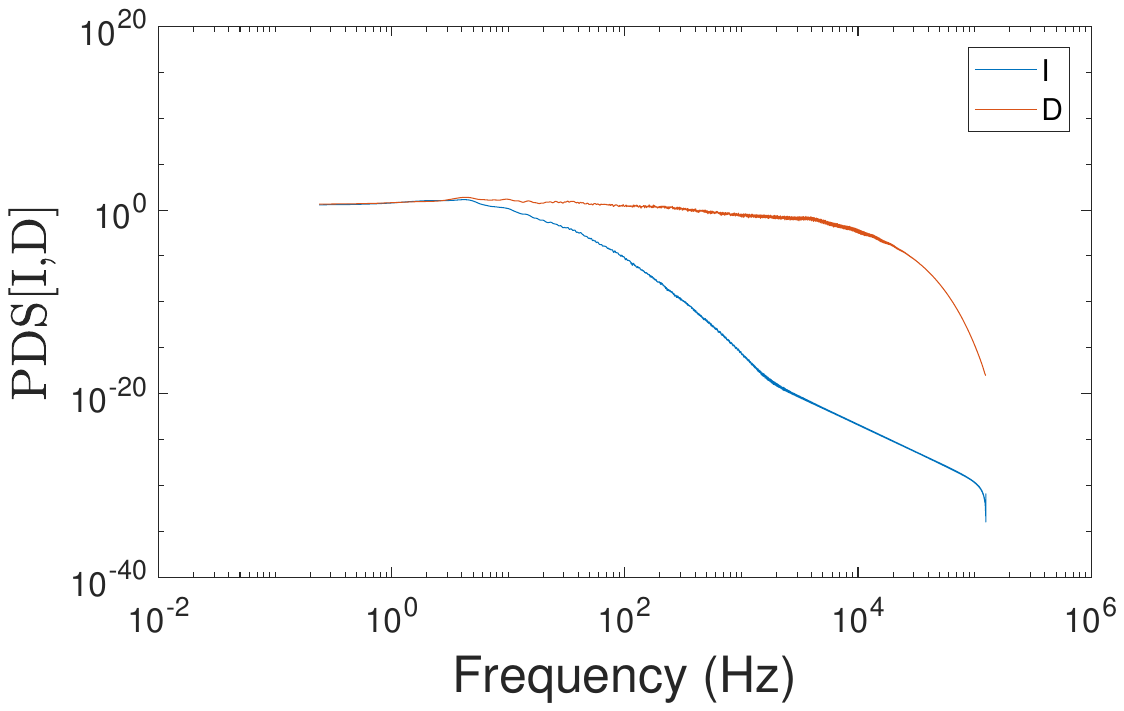}
}
\vspace{-4cm}
\caption{Power Density Spectral (PDS) of the injected (blue) and dissipated (red) power in the GOY--shell model with N=20, $\nu= 7.62 \,10^{-7}$ and $f_0=0.071$}
\label{specIDGOY}
\end{figure}

 The injected power $I=f_0(u_{4,r}+ u_{4,i})$ and the dissipated power $D=\sum_{n=1}^{20}\nu k_n |u^2|$ show very different temporal traces, the latter being very intermittent as shown in figure \ref{TrcTpesIDGOY}. 
Nevertheless, it is clear in figure \ref{specIDGOY} that the PDS at vanishing frequency of injected and dissipated power converge to the same value as expected from equation (\ref{PDSstat}).

\begin{figure}
\vspace{-3cm}
\resizebox{.5\textwidth}{!}{%
\includegraphics{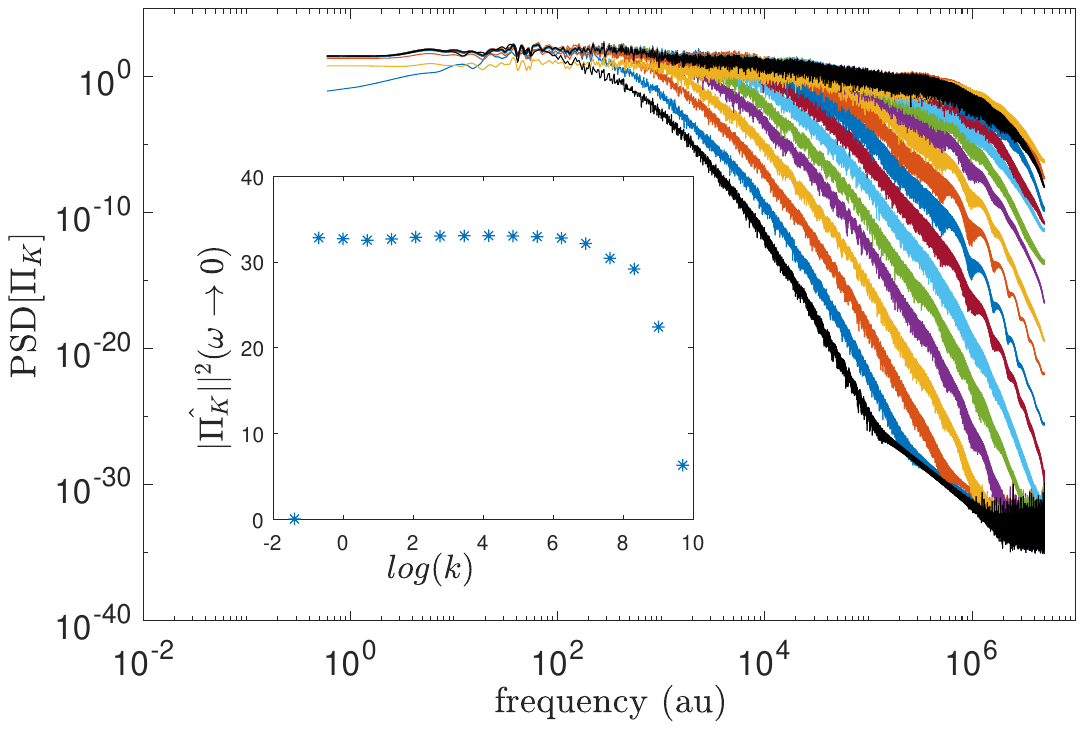} 
}
\vspace{-4cm}
\caption{Power Density Spectra (PDS) of the rate of energy transfer $\Pi_K$ for the all shell layers. Each color represents the PDS of the energy flux at a given shell, the high frequency part increasing with the wavenumber $K$. The black curve corresponds to the PDS of the injected power.
The blue asterix in inset show the smallest frequency limits of this PDS, $|\hat{\Pi}_K|^2(\omega\rightarrow 0)$.}
\label{PSDPiK} 
\end{figure}

Moreover, for the GOY-shell model, another energy balance can be written for the truncated energy $E_M=\sum_{n=1}^M|u_n^2|/2$ with $4<M<20$. One gets

\begin{equation}
\frac{d E_M}{dt}=I-\Pi_K',
\label{GOYdyn1}
\end{equation}
where $\Pi_K'=\Pi_K+\sum_{n=0}^M\nu k_n |u^2|$ with 
\begin{equation}
\Pi_K=jK\left(u_{M+2}u_{M+1}u_{M}+u_{M+1}u_{M}u_{M-1}/4\right)^*+cc
\label{DefPiK} 
\end{equation}
is the energy flux through the shell $M$ and $K=2^M/16$. In the inertial range i.e. for $K\ll(\langle I \rangle /\nu^3)^{1/4}$, the term $\sum_{n=1}^M\nu k_n |u^2|$ due to dissipation, is negligible in the expression of $ \Pi_K'$. Because the balance (\ref{GOYdyn1}) is formally equivalent to equation (\ref{EBalance}), one deduces that in the inertial range
\begin{eqnarray}
\label{PiBalancea}
\lim_{\omega\rightarrow 0} \left[|\widehat{I}(\omega)|^2\right] =& \lim_{\omega\rightarrow 0} \left [|\widehat{\Pi}_K(\omega)|^2\right]\\
\label{PiBalance}
\tau_I\sigma(I)^2=&\tau_{\Pi_K}\sigma(\Pi_K)^2
\end{eqnarray}
This is indeed the case for the GOY--shell model as shown in figure \ref{PSDPiK} where we plot the PDS of $|\widehat{\Pi}_K(\omega)|^2$ for each $K=2^M/16$ with $2 \leq M \leq18$. The inset shows a large inertial range of scales where the limit at vanishing frequency is constant. It falls down at the largest wave numbers, $K$, because the viscous damping must be taken into account in the dissipative range.

Finally we can compute separately the variance of $\Pi_K$ as function of $K$ and deduce the scaling of the correlation time $\tau_{\Pi_K}= \frac{1} {\sigma(\Pi_K)^2}\;\lim_{\omega\rightarrow 0} \left [|\widehat{\Pi_K}(\omega)|^2\right]$.

\begin{figure}
\centerline{\resizebox{0.57\textwidth}{!}{%
\includegraphics{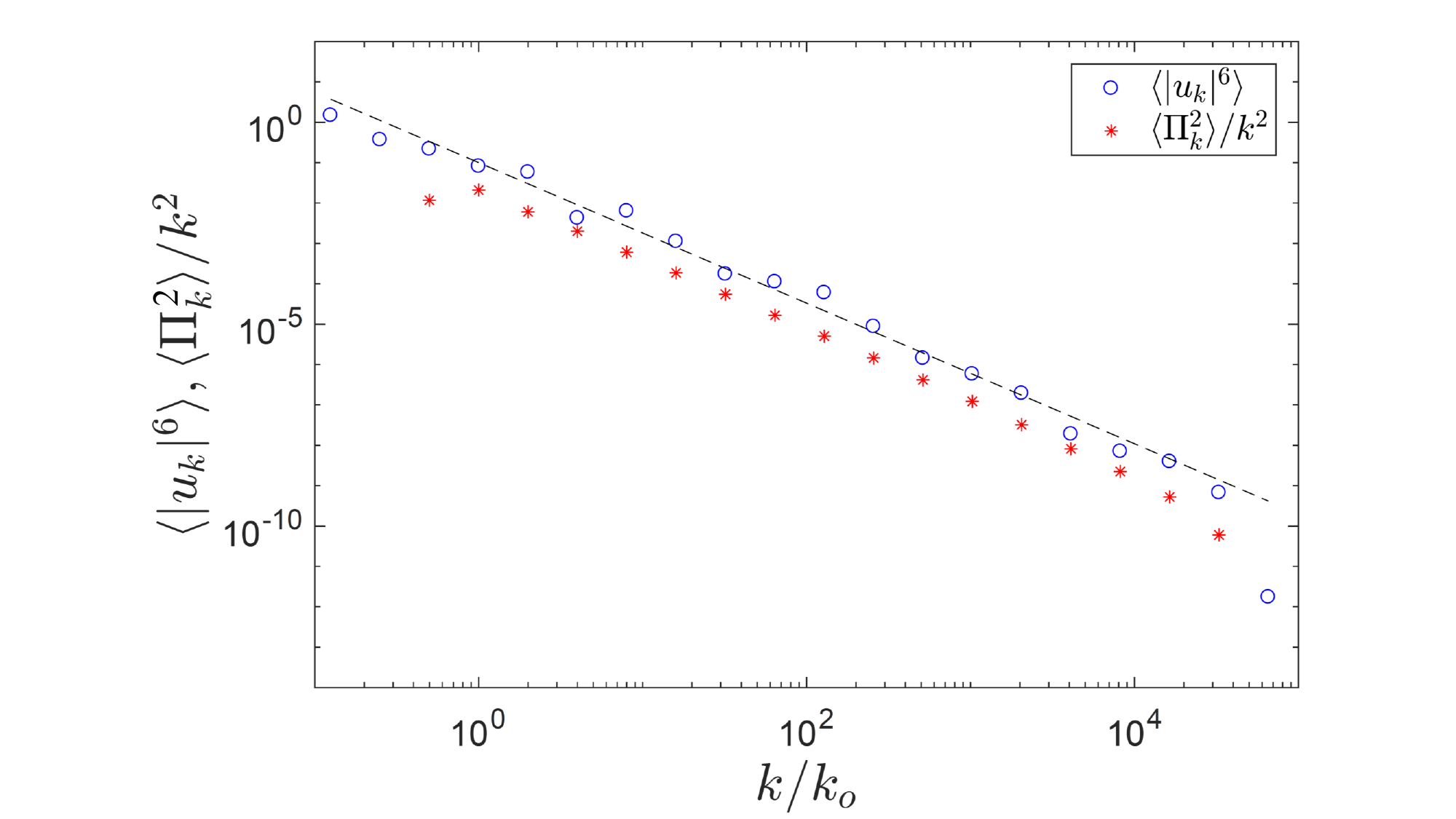}}
}
\caption{Comparison between the scaling of $\langle |u_M|^6\rangle$ and $\sigma(\Pi_K)^2/K^2$. The black dashed line is the prediction of the SL model and the black dot-dashed line (almost mingled) is obtained from a fit between the 5th and 16th shells of the GOY--shell model.}
\label{siPi} 
\end{figure}

Figure \ref{siPi} shows that $\sigma(\Pi_K)^2\propto \langle |u_M|^6\rangle K^2 $ in agreement with a simple dimensional estimate. We also get $ \langle |u_M|^6\rangle\propto K^{-\zeta(6)}$ with $\zeta(6)=1.73\pm 0.01$ from a fit between the 5th shell and the 16th shell, in perfect agreement with the SL model (for which $\zeta(6)=1.737$). All together, \begin{equation}
\sigma(\Pi_K)^2\propto \langle |u_M|^6\rangle K^2\propto \frac{U^6}{L^2} (KL)^{-\zeta(6)+2} \propto K^{0.27}.
\label{sigmapi}
\end{equation}
Note that this scaling law for the variance of the fluctuations of the energy flux is in reasonable agreement with direct numerical simulations of hydrodynamic turbulence \cite{Aoyama}. Since the left hand side of equation (\ref{PiBalance}) is independent of $K$, it implies $\tau_{\Pi_K}\propto K^{-0.27}$.

This does not correspond to the characteristic turnover time at each scale that is of order $\tau_M\sim\langle |u_M|^{-1} \rangle K^{-1}\propto (L/U) (KL)^{-\zeta(-1)-1}$. Indeed that gives $\tau_M\propto K^{-0.6}$ with the SL model. Nevertheless, there is no reason that the correlation time of the coarse-grained quantities involved in equation (\ref{GOYdyn1}) reduces the turnover time of the smallest scale implied in the coarse graining process.

To tackle the scaling of $\tau_{\Pi_K}$, one assumes that there exists a single scaling in $K$ for all quantities coarse-grained up to the wave number $K$, especially those involved in (\ref{GOYdyn1}). In order to find this scaling law, we focus on the coarse-grained velocity $U_K=\sum_{i=1}^M u_i$. By definition, one has
\begin{equation}
\label{TauPiscal1}
\tau_{U_K}= \frac{\int_0^\infty \langle U_K(t)\cdot U_K(t+\tau)\rangle d\tau}{\sigma(U_K)^2}
\end{equation}
where $\tau_{U_K}$, the correlation time of $U_K$, shares the same scaling in $K$ that $\tau_{\Pi_K}$ by assumption. The computation of $\tau_{U_K}$ can be pushed forward:

\begin{eqnarray}
\label{TauPiscal2}
\tau_{U_K}\sim &\frac{\sum_{n=1}^M\int_0^\infty \langle u_n(t)\cdot u_n(t+\tau)\rangle d\tau}{\sum_{n=1}^M\langle|u_n|^2\rangle}\\
\label{TauPiscal3}
\tau_{U_K}\sim &\frac{\sum_{n=1}^M\tau_n\sigma(u_n)^2}{\sum_{n=1}^M\langle|u_n|^2\rangle}
\end{eqnarray}
where we take into account the short range of the interaction to neglect the cross terms in equation (\ref{TauPiscal2}) (assuming a random phase between the different complex velocity components $u_n$) and where dimensional analysis imposes $\tau_n\sim\langle |u_n|^{-1} \rangle k_n^{-1}$. 

Focusing on the inertial range where the scaling laws apply and taking the continuous limit for the wave vectors, $\tau_{U_K}$ can be approximated by
\begin{equation}
\tau_{U_K}\propto\left(\frac{1-\zeta(2)}{-\zeta(-1)-\zeta(2)}\right)\frac{K^{-\zeta(-1)-\zeta(2)}-1}{K^{1-\zeta(2)}-1}
\label{Taupi}
\end{equation}
The limit at large $K$ depends on the value of $\zeta(2)$ and $\zeta(-1)$. From our simulation of the GOY--shell model, one gets $\zeta(2)=0.72\pm0.02$ and $\zeta(-1)=-0.45\pm0.02$ between the 5th shell and the 16th shell, in good agreement with the prediction of She-Leveque ( $\zeta(2)=0.703$ and $\zeta(-1)=-0.422$). 
Hence in the limit $K >> 1$ and with our assumption of a similar scaling for all the coarse-grained variables, one deduces from equation (\ref{Taupi}) that
\begin{equation}
\tau_{U_K}\propto\tau_{\Pi_K}\propto \frac{L}{U}(K L)^{\zeta(2)-1}
\label{Taupi2}
\end{equation}
i.e. $\tau_{\Pi_K}\propto K ^{-0.28}$ in good agreement with the scaling of $\sigma(\Pi_K)^2$. Note that this final scaling law does not depend on the precise definition of $\tau_n$. For instance using $\tau_n=\langle |u_n| \rangle^{-1} k_n^{-1}$ (replacing $\zeta(-1)$ by $-\zeta(1)$ with $\zeta(1)=0.37$) would have led to the same scaling for $\tau_{\Pi_K}$.

Interestingly, using equations (\ref{sigmapi}) and (\ref{Taupi2}) and taking into account the constraint (\ref{PiBalance}), imposes the relation
\begin{equation}
\zeta(6)-\zeta(2)-1=0.
\label{TauPiscal}
\end{equation}
The simulation of the GOY-shell model gave: $\zeta(2)=0.72\pm0.02$, $\zeta(6)=1.73\pm0.02$. Hence relation (\ref{TauPiscal}) holds within a percent. Notice that the K41 theory gives $\zeta(2)=2/3$ and $\zeta(6)=2$ which is less satisfactory \cite{NoteK41}. 

In the case of the log-normal model, equation (\ref{TauPiscal}) fixes the intermittency parameter $\mu$ to 3/10 which is compatible with the values extracted from numerical and experimental data \cite{Sreenivasan1993,Wang1996}.\\


To sum up, we have shown that some deviations from the K41 prediction $\zeta(p)=p/3$ are required to satisfy the constraints on the power fluctuations imposed by stationarity in the GOY--shell model. These required deviations are also a trace of the model intermittency. The intermittent scaling exponents indeed satisfy equation (\ref{TauPiscal}) imposed by the second order stationary balance (\ref{PiBalance}).

We want to conclude this discussion with the second order stationary relation between injected and dissipated power (\ref{PDSstat}--\ref{BalanceII}). Although the variance of the injected power and the one of the dissipated power strongly differ in the GOY--Shell model (see the temporal traces in figure \ref{TrcTpesIDGOY} and the PDS figure \ref{specIDGOY}), the relation (\ref{BalanceII}) holds. Since the left hand side of this relation is independent of the Reynolds number, so does the right hand side. We can expect that the variance and the correlation time of the dissipation depend on the Reynolds number, but their product must not. A systematic analysis of the variance and correlation time of $D$ in the GOY-shell model for different viscosities would be interesting but we postpone it to further studies. 

This previous remark emphasises an important difference between shell models and turbulent flows. 
 In the later case, the dissipated power is a global quantity averaged over the entire flow volume. Its temporal trace is much less intermittent than in figure \ref{TrcTpesIDGOY} (bottom) because the average over volume smears out incoherent small scale dissipative bursts. Direct Numerical Simulations (DNS) indeed show that the fluctuations of the dissipation almost follow those of the injected power \cite{Pumir,Cardesa}. Both are only affected by  
 structures existing on an integral time scale. Actually (\ref{BalanceII}) holds for real 3D turbulence because both $\sigma(I)\sim\sigma(D)$ and $\tau_I\sim\tau_D$.  
 
 Nevertheless, the reasoning presented here on the GOY--shell model, might also apply to a coarse-grained fluid element that we follow in a Lagrangian way in 3D turbulent flow. Indeed, in this case, the  energy balance contains only two terms, an injection one implying the large scale forces, including the pressure work, applied on the coarse-grained fluid particles, and a dissipative term containing the viscous dissipation and the energy flux through the smaller scales than the one of  coarse-graining. The same dimensional arguments can be  applied to this last term. Using some ergodic hypothesis, this may explain why the relation between scaling exponents established previously is compatible with the one measured in Eulerian turbulent flows   \cite{AnselmetGagneHoppfinger,ESS,Aoyama}. We hope this will motivate further direct numerical simulations to evidence constraints on the fluctuations of the various terms involved in the local energy budget \cite{Vassilicos}.

\

\end{document}